\documentclass[%
superscriptaddress,
nofootinbib,
nobibnotes,
amsmath,amssymb,
aps,
floatfix]{revtex4-2}

\usepackage{placeins}
\usepackage{booktabs}
\usepackage{dcolumn}
\usepackage{array}
\usepackage{longtable}
\usepackage{float}
\usepackage{hyphenat}
\usepackage{natbib}
\usepackage{hyperref}

\usepackage{amsmath,amsfonts,amsthm,amssymb}
\usepackage{bm,graphicx,graphics,color}
\usepackage{epsf,epsfig}
\usepackage[all]{xy}
\newcommand*{\B}[1]{\ifmmode\bm{#1}\else\textbf{#1}\fi}

\def\bx{\mathbf{x}}

\def\bv{\mathbf{v}}
\def\bV{\mathbf{V}}

\def\no{\nonumber}
\def\lb{\label}
\def\be{\begin{equation}}
\def\ee#1{\label{#1}\end{equation}}
\newcommand{\ben}{\begin{eqnarray}}
\newcommand{\een}{\end{eqnarray}}
\usepackage[english]{babel}

\usepackage[letterpaper,top=2cm,bottom=2cm,left=3cm,right=3cm,marginparwidth=1.75cm]{geometry}

\usepackage{amsmath}
\usepackage{graphicx}
\begin{document}

\title{Relaxation-Time Model for the Post-Newtonian Boltzmann Equation}

\author{Gilberto M. Kremer}
\email{kremer@fisica.ufpr.br}
\affiliation{Departamento de F\'{i}sica, Universidade Federal do Paran\'{a}, Curitiba 81531-980, Brazil}

\begin{abstract}
The non-equilibrium contributions to the post-Newtonian hydrodynamic equations are determined from a relaxation-time model of the post-Newtonian Boltzmann equation. The Chapman-Enskog method is used to calculate the non-equilibrium distribution function. The components of the energy-momentum tensor are found from the knowledge of the  non-equilibrium and the post-Newtonian equilibrium Maxwell-J\"uttner distribution functions. The linearized field equations for the  mass, momentum and internal energy densities coupled with the three Poisson equations of the post-Newtonian approximation are investigated by considering a plane wave representation of the fields. The constitutive equations for the viscous stress and heat flux vector are obtained and it is shown that the transport coefficients of  shear viscosity and heat conductivity do  depend on the Newtonian gravitational potential.
\end{abstract}


\maketitle

\section{Introduction}
\lb{s.1}
 In the seminal work of Einstein, Infeld and Hoffman \cite{Eins} it was proposed a method of successive approximations in powers of $1/c^2$ for the solution of  Einstein's field equations, which become the basis of the post-Newtonian approximation for the determination of  the  energy-momentum tensor components as well as the Eulerian hydrodynamic equations in the first \cite{Ch1,Wein} and second \cite{ChNu} approximations.

 The post-Newtonian version of the Boltzmann equation in the first and in the second approximations  were determined in \cite{Ped,Rez}  and \cite{PGMK,GGKK}, respectively. In \cite{PGMK,GGKK} the energy-momentum tensor components were obtained from the equilibrium Maxwell-J\"uttner distribution function \cite{KRW} in the first and second post-Newtonian approximations and the Eulerian hydrodynamic equations from a collisionless post-Newtonian Boltzmann equation were determined.

The inclusion of non-equilibrium terms in the post-Newtonian theory was investigated in \cite{GR,JH} within the framework of a phenomenological theory of a viscous, heat conducting and compressible fluid. On the other hand, the inclusion of  non-equilibrium terms  in the hydrodynamic equations which follow from the post-Newtonian Boltzmann equation was considered in \cite{GKN}. In this work  the  hydrodynamic equations resulted from a post-Newtonian Maxwell-Enskog transfer equation together with a post-Newtonian Grad's distribution function which takes into account   the non-equilibrium fields of viscous stress and  heat flux vector.  

One interesting subject to be investigate is the determination of  the post-Newtonian hydrodynamic equations for a viscous and heat conducting fluid from the post-Newtonian Boltzmann equation where the particle collisions are taken into account through the collision operator of the Boltzmann equation. Here we shall adopt a relaxation-time model for the collision operator which is known in the non-relativistic framework as the Bhatnagar-Gross-Krook (BGK) model (see e.g. \cite{Chap,GK}) and in the relativistic one as the Marle model \cite{Mar,CK}.

We use the Chapman-Enskog method to determine the non-equilibrium distribution function from the post-Newtonian BGK (Marle) model of the Boltzmann equation and the post-Newtonian Maxwell-J\"uttner distribution function. From the knowledge of the non-equilibrium distribution function the non-equilibrium contributions to the energy-momentum tensor are calculated.  The  linearized field equations for the 
 mass, momentum and internal energy densities are determined from the  particle four-flow and energy-momentum tensor conservation laws. These linearized field equations are coupled with three Poisson equations from the post-Newtonian approximation and a solution of the coupled system of equations is found in terms of a plane wave representation of the fields. Furthermore, the constitutive equations for the viscous stress and heat flux vector -- which correspond to the Navier-Stokes and Fourier laws, respectively -- are obtained from the Eckart decomposition \cite{Eck} of the energy-momentum tensor. It is shown that the transport coefficients of  shear viscosity and heat conductivity do  depend on the Newtonian gravitational potential.

 The paper is structured as follows: in Section \ref{sec2} we introduce the relaxation-time model of the post-Newtonian Boltzmann equation and determine the non-equilibrium distribution function.  The particle four-flow and the energy-momentum tensor components are calculated on the basis of the equilibrium Maxwell-J\"uttner and non-equilibrium distribution functions in Section \ref{sec3}. The linearized field equations are determined in Section \ref{sec4} and a plane wave solution of the linearized field equations coupled with the three Poisson equations of the post-Newtonian approximation is analyzed. In Section \ref{sec5} the constitutive equations for the viscous stress and heat flux vector are obtained and the transport coefficients of shear viscosity and thermal conductivity are identified. In the last section the conclusions of the work are stated.

\section{Relaxation-Time Model}\lb{sec2}

In  the phase space spanned by the space coordinates $\bx$ and velocity of the particles $\bv$  a state of a monatomic gas   is characterized  by the one-particle distribution function $f(\bx, \bv, t)$ and its spacetime evolution is governed by Boltzmann equation. In the first post-Newtonian approximation the Boltzmann equation is given by \cite{Ped,PGMK,GGKK}
\ben\no
&&\bigg[\frac{\partial f}{\partial t}+v_i\frac{\partial f}{\partial x^i}+\frac{\partial f}{\partial v^i}\frac{\partial U}{\partial x^i}\bigg]\bigg[1+\frac1{c^2}\left(\frac{v^2}2+U\right)\bigg]+\frac1{c^2}\frac{\partial f}{\partial v^i}\bigg\{v_j\bigg(\frac{\partial\Pi_i}{\partial x^j}-\frac{\partial\Pi_j}{\partial x^i}\bigg)
\\\lb{tr8a}
&&
-3v_i\frac{\partial U}{\partial t}+\frac{\partial\Pi_i}{\partial t}
+2\frac{\partial\Phi}{\partial x^i}-4U\frac{\partial U}{\partial x^i}-4v_iv_j\frac{\partial U}{\partial x^j}+v^2\frac{\partial U}{\partial x^i}
\bigg\}=\mathcal{Q}(f,f).
\een
Here $\mathcal{Q}(f,f)$ denotes the collision operator of the Boltzmann equation which takes into account the binary collisions of the particles and refers to an integral of the product of two particle distribution functions at collision. Furthermore, the Newtonian gravitational potential $U$, the scalar gravitational potential  $\Phi$ and the vector gravitational potential  $\Pi_i$  satisfy  Poisson equations,  which are obtained from the first post-Newtonian approximation  of Einstein's field equations and read \cite{Ch1,GGKK}
\ben\lb{tr4a}
&&\nabla^2U=-4\pi G\rho,
\qquad
\nabla^2\Phi=-4\pi G\rho\left(V^2+U+\frac\varepsilon2+\frac{3p}{2\rho}\right),
\\\lb{tr4c}
&&\nabla^2\Pi_i=-16\pi G\rho V_i+\frac{\partial^2U}{\partial t\partial x^i}.
\een
Above  $\bV$ denotes the  hydrodynamic three-velocity, $G$ the universal gravitational constant and $\varepsilon, p$ the specific internal energy and hydrostatic pressure of the gas, respectively.  The gauge condition $3{\partial U}/{\partial t}+{\partial \Pi_i}/{\partial x^i}=0$ for the gravitational potentials $U$ and $\Pi_i$ holds.
%In  the components of the metric tensor are given by \cite{Ch1,Wein,GGKK}:
%\ben\lb{tr3a} g_{00}=1-\frac{2U}{c^2}+\frac2{c^4}\left(U^2-2\Phi\right),\quad&& g_{0i}=\frac{\Pi_i}{c^3},\quad g_{ij}=-\left(1+\frac{2U}{c^2}\right)\delta_{ij},\een  

  In the BGK (Marle) model the collision operator is replaced by the  difference between the  one-particle distribution function and its  equilibrium value multiplied by a frequency $\nu$ which is of order of the collision frequency. 

 The  one-particle distribution function at equilibrium is determined from  the relativistic Boltzmann equation by considering that the collision operator vanishes at equilibrium. In the relativistic theory the equilibrium distribution function is the Maxwell-J\"uttner distribution function (see e.g \cite{CK}) and its first post-Newtonian approximation was determined in \cite{KRW} and reads
\ben\lb{tr9a}
f_{MJ}=f_0
\Bigg\{ 1-\frac{1}{c^2}\bigg[\frac{15 kT}{8 m}+\frac{m (V_i\mathcal{V}_i)^2}{2 kT}+\frac{2 m U \mathcal{V}^2}{kT}
+\frac{3 m \mathcal{V}^4}{8 kT}+\frac{m V^2 \mathcal{V}^2}{2 kT}+\frac{m (V_i\mathcal{V}_i) \mathcal{V}^2}{kT}\bigg]\Bigg\},
\een
where $f_0$ denotes the non-relativistic Maxwellian distribution function, namely
\ben
f_0=\frac{\rho\, e^{-\frac{m\mathcal{V}^2}{2kT}}}{(2\pi m^\frac53kT)^{\frac32}}.
\een
In the above equation $\rho$ is the mass density, $T$ the absolute temperature, $m$ the rest mass of a gas particle and $k$ the Boltzmann constant. Furthermore,  $\mathcal{V}_i=v_i-V_i$ is the so-called peculiar velocity which is the difference of the particle velocity $v_i$ and the hydrodynamic velocity $V_i$.

By considering that the relativistic  equilibrium distribution function is the Maxwell-J\"uttner distribution $f_{MJ}$,  the collision operator is written as 
\ben\lb{tr8b}
\mathcal{Q}(f,f)=-\nu(f-f_{MJ})=-\nu f_{NE},
\een
where $f_{NE}$ is the non-equilibrium distribution function.

 For the determination of the non-equilibrium distribution function we shall rely on the Chapman-Enskog method (see e.g. \cite{Chap,GK} and insert the equilibrium Maxwell-J\"uttner distribution function (\ref{tr9a}) into the left-hand side of the Boltzmann equation (\ref{tr8a})  and compute the non-equilibrium distribution function by considering the BGK (Marle) model (\ref{tr8b}). Hence it follows
\ben\no
&&\bigg[1+\frac1{c^2}\left(\frac{v^2}2+U\right)\bigg]\bigg[\frac{\partial f_{MJ}}{\partial \rho}\bigg(\frac{d\rho}{dt}+\mathcal{V}_i\frac{\partial \rho}{\partial x^i}\bigg)+\frac{\partial f_{MJ}}{\partial T}\bigg(\frac{dT}{dt}+\mathcal{V}_i\frac{\partial T}{\partial x^i}\bigg)+\frac{\partial f_{MJ}}{\partial V_i}\bigg(\frac{dV_i}{dt}+\mathcal{V}_j\frac{\partial V_i}{\partial x^j}\bigg)
\\\no
&&
\qquad+\frac{\partial f_{MJ}}{\partial U}\bigg(\frac{dU}{dt}+\mathcal{V}_i\frac{\partial U}{\partial x^i}\bigg)+\frac{\partial f_{MJ}}{\partial v^i}\frac{\partial U}{\partial x^i}\bigg]
+\frac1{c^2}\frac{\partial f_{MJ}}{\partial v^i}\bigg\{v_j\bigg(\frac{\partial\Pi_i}{\partial x^j}
-\frac{\partial\Pi_j}{\partial x^i}\bigg)-3v_i\frac{\partial U}{\partial t}+\frac{\partial\Pi_i}{\partial t}
\\\lb{tr10a}
&&\qquad+2\frac{\partial\Phi}{\partial x^i}-4U\frac{\partial U}{\partial x^i}-4v_iv_j\frac{\partial U}{\partial x^j}+v^2\frac{\partial U}{\partial x^i}
\bigg\}=-\nu f_{NE},
\een
where $d/dt=\partial/\partial t+V_i\partial/\partial x^i$ denotes the material time derivative and
%\begin{widetext}
\ben\lb{tr11a}
&&\frac{\partial f_{MJ}}{\partial \rho}=\frac{ f_{MJ}}{\rho},\qquad \frac{\partial f_{MJ}}{\partial U}=-f_0\frac{2m\mathcal{V}^2}{kTc^2},
\\\no
&&\frac{\partial f_{MJ}}{\partial T}=\frac{f_0}{T}\bigg\{\frac{m\mathcal{V}^2}{2kT}-\frac32+\frac{1}{c^2}\bigg[\frac{15kT}{16m}\bigg(1-\frac{m\mathcal{V}^2}{kT}+\frac{m^2\mathcal{V}^4}{k^2T^2}\bigg)+\frac{5m}{2kT}\bigg(2U\mathcal{V}^2+\frac{ (V_i\mathcal{V}_i)^2}{2}
\\\lb{tr11b}
&&+\frac{V^2\mathcal{V}^2}{2}
+(V_i\mathcal{V}_i)\mathcal{V}^2\bigg)-\frac{m^2}{2k^2T^2}\bigg(2U\mathcal{V}^4
+\frac{ (V_i\mathcal{V}_i)^2\mathcal{V}^2}{2}+\frac{V^2\mathcal{V}^4}{2}+(V_i\mathcal{V}_i)\mathcal{V}^4+\frac{3\mathcal{V}^6}8\bigg)
\bigg]\bigg\},\\\no
&&\frac{\partial f_{MJ}}{\partial V^i}=\frac{mf_0}{kT}\bigg\{\mathcal{V}_i+\frac{1}{c^2}\bigg[4U\mathcal{V}_i\bigg(1-\frac{m\mathcal{V}^2}{2kT}\bigg)
-\frac{15kT}{8m}\mathcal{V}_i+(\mathcal{V}_j V_j)V_i+\mathcal{V}_iV^2\bigg(1-\frac{m\mathcal{V}^2}{2kT}\bigg)
\\\lb{tr11c}
&&+(\mathcal{V}_j V_j)\mathcal{V}_i\bigg(1-\frac{m\mathcal{V}^2}{kT}\bigg)+\frac{\mathcal{V}_i\mathcal{V}^2}2\bigg(1-\frac{3m\mathcal{V}^2}{4kT}\bigg)
-\frac{m(\mathcal{V}_j V_j)^2}{2kT}\mathcal{V}_i\bigg]\bigg\},
\\\no
&&\frac{\partial f_{MJ}}{\partial v^i}=-\frac{mf_0}{kT}\bigg\{\mathcal{V}_i+\frac{1}{c^2}\bigg[4U\mathcal{V}_i\bigg(1-\frac{m\mathcal{V}^2}{2kT}\bigg)+V_i(\mathcal{V}^2+\mathcal{V}_j V_j)-\frac{15kT}{8m}\mathcal{V}_i
\\\lb{tr11c}
&&
+\mathcal{V}_i\bigg(V^2+2\mathcal{V}_j V_j+\frac{3\mathcal{V}^2}{2}\bigg)
-\frac{m\mathcal{V}_i}{kT}\bigg(\frac{(\mathcal{V}_j V_j)^2}{2}
+\frac{V^2\mathcal{V}^2}2+(\mathcal{V}_jV_j)\mathcal{V}^2+\frac{3\mathcal{V}^4}8\bigg)\bigg]\bigg\}.
\een
%\end{widetext}

As usual in the Chapman-Enskog method the material time derivatives are eliminated from the non-equilibrium distribution function by using the Eulerian balance equations for the mass density $\rho$, hydrodynamic velocity $V_i$ and absolute temperature $T$. 

The Eulerian mass density and the momentum density balance equations in the first post-Newtonian approximation  are \cite{Ch1,GGKK}
\ben\lb{tr12a}
\frac{d\rho\left[1+\frac1{c^2}\left(\frac{V^2}2+3U\right)\right]}{d t}+ \rho\left[1+\frac1{c^2}\left(\frac{V^2}2+3U\right)\right]\frac{\partial V_i}{\partial x^i}=0,
\\\no
\rho\frac{dV_i}{dt}+\frac{\partial p}{\partial x^i}\left[1-
\frac1{c^2}\left(V^2+4U+\varepsilon+\frac{p}\rho\right)\right]
-\rho\frac{\partial U}{\partial x^i}\bigg[1+\frac1{c^2}(V^2-4U)\bigg]
\\\lb{tr12b}
+\frac\rho{c^2}\bigg[\bigg(\frac1\rho\frac{\partial p}{\partial t}-\frac{\partial U}{\partial t}
+4\frac{dU}{dt}\bigg)V_i-2\frac{\partial \Phi}{\partial x^i}-\frac{d\Pi_i}{dt}+V_j\frac{\partial \Pi_j}{\partial x^i}
\bigg]=0.
\een

For the determination of the Eulerian internal energy density  balance equation $\rho\varepsilon$  in the  first post-Newtonian approximation one has to go to the second post-Newtonian approximation, since within the framework of the first post-Newtonian approximation one recover only its Newtonian expression. The  Eulerian internal energy density  balance equation reads\footnote{This equation corrects some misprints in \cite{GGKK,PGMK}}
\be
 \frac{d\varepsilon}{dt}+\frac{p}\rho\frac{\partial V_i}{\partial x^i} +\frac{3p}{\rho c^2}\frac{dU}{dt}
+\frac{ pV_i}{\rho c^2}\bigg[\frac{\partial U}{\partial x^i}-\frac1\rho\frac{\partial p}{\partial x^i}\bigg]
 =0.
 \ee{tr12c}
From the above equation follows the expression for the material time derivative of the absolute temperature, if we take into account the relationship for  the specific internal energy in the first post-Newtonian approximation which comes from the relativistic kinetic theory of gases (see e.g. \cite{CK})
\ben\lb{tr12d}
\varepsilon=\frac{3kT}{2m}\left(1+\frac{5kT}{4mc^2}\right).
\een

 \section{Particle four-flow and energy-momentum tensor components}\lb{sec3}

In  the relativistic kinetic theory of gases the  particle four-flow $N^\mu$ and the energy-momentum tensor $T^{\mu\nu}$ are given in terms of the one-particle distribution function $f(\bx,\bv,t)$ by \cite{CK,GGKK}
\ben\lb{tr13a}
N^\mu=m^4c\int u^\mu f\frac{\sqrt{-g}\,d^3 u}{u_0},\qquad
T^{\mu\nu}=m^4c\int u^\mu u^\nu f\frac{\sqrt{-g}\,d^3 u}{u_0}.
\een
Here     $u^\mu=p^\mu/m$  (with $u^\mu u_\mu=c^2$) denotes the gas particle four-velocity whose components in the first post-Newtonian approximation read \cite{Ch1,Wein,GGKK}
\ben\lb{tr13b}
u^0=c\bigg[1+\frac1{c^2}\bigg(\frac{v^2}2+U\bigg)\bigg],\quad u^i=v_i \frac{u^0}c,
\een
where $\bv$ is the particle three-velocity. Furthermore,  ${\sqrt{-g}\,d^3 u}/{u_0}$ is an invariant integration element whose  first post-Newtonian approximation  was determined in \cite{KRW} and is given by
\ben\lb{tr13c}
\frac{\sqrt{-g}\, d^3 u}{u_0}=
\left\{1+\frac1{c^2}\left[2v^2+6U\right]\right\}\frac{d^3 v}c.
\een

Once the one-particle distribution function $f=f_{MJ}+f_{NE}$ and the invariant integration element are known, one can determine the components of the  particle four-flow $N^\mu$ and energy-momentum tensor $T^{\mu\nu}$. Indeed, if we insert (\ref{tr9a}), (\ref{tr10a}), (\ref{tr13b}) and (\ref{tr13c}) into (\ref{tr13a}) and integrate the resulting equations we get
\ben\lb{tr14a}
&&N^0=\frac{\rho c}m\left[1+\frac1{c^2}\left(\frac{V^2}2+U\right)\right],\qquad N^i=N^0\frac{V_i}c,
\\\lb{tr14b}
&&T^{00}= \rho c^2\bigg[1+\frac1{c^2}\bigg(V^2+\frac{3kT}{2m}+2U\bigg)+\mathcal{O}(c^{-4})\bigg],
\\\lb{tr14c}
&&T^{i0}= c\rho V_i\bigg[1+\frac1{c^2}\bigg(V^2+2U+\frac{5kT}{2m}\bigg)\bigg]+T^{i0}_{NE},
\\\lb{tr14d}
&&T^{ij}=\rho V_iV_j\left[1+\frac1{c^2}\left(V^2+2U+\frac{5kT}{2m}\right)\right]
+p\left(1-\frac{2U}{c^2}\right)\delta_{ij}+T^{ij}_{NE}.
\een
Note that there are no non-equilibrium contributions to the components of the particle four-flow (\ref{tr14a}.
The  non-equilibrium contribution to $T^{00}$ is of order $\mathcal{O}(c^{-4})$ (the order of the $n$th inverse power of light speed is denoted by $\mathcal{O}(c^{-n})$) while the non-equilibrium contributions to the energy-momentum tensor components $T^{0i}_{NE}$ and $T^{ij}_{NE}$ are associate with terms related with the collision frequency $\nu$ and read
\ben\lb{tr14e}
&&T^{i0}_{NE}=\frac{-p}{\nu c}\bigg[\frac{5k}{2m}\frac{\partial T}{\partial x^i}+\Delta_{ijkl}V_j\frac{\partial V_k}{\partial x^l}\bigg],
\\\no
&&T^{ij}_{NE}=-\frac{p}\nu\bigg\{\bigg[1+\frac1{c^2}\bigg(\frac{5kT}{2m}-U+\frac{V^2}2\bigg)\bigg]\Delta_{ijkl}V_j\frac{\partial V_k}{\partial x^l}+\frac{1}{ c^2}\Delta_{ijkl}V_k\left(\frac{\partial U}{\partial x^l}-\frac1\rho\frac{\partial p}{\partial x^l}\right)
\\\lb{tr14f}
&&
-\frac2{3c^2}V_iV_j\frac{\partial V_k}{\partial x^k}+\frac1{c^2}\bigg(V_j\frac{\partial}{\partial x^i}+V_i\frac{\partial}{\partial x^j}\bigg)
\bigg[\frac{5kT}{2m}+\frac{V^2}2\bigg]
\bigg\}.
\een
%\end{widetext}
Here we have introduced the fourth-order tensor
\ben
\Delta_{ijkl}=\delta_{ik}\delta_{jl}+\delta_{il}\delta_{jk}-\frac23\delta_{ij}\delta_{kl}.
\een

\section{Linearized Field equations }\lb{sec4}

The thermodynamic theory of a  single relativistic fluid is described by the fields of particle four-flow $N^\mu$ and energy-momentum tensor $T^{\mu\nu}$ where their  
hydrodynamic equations follow from the conservation laws
\ben\lb{tr15}
{N^{\mu}}_{;\mu}=\frac{\partial N^\mu}{\partial x^\mu}+{\Gamma^\mu}_{\mu\sigma}N^\sigma=0, \qquad {T^{\mu\nu}}_{;\nu}=\frac{\partial T^{\mu\nu}}{\partial x^\nu}+{\Gamma^\mu}_{\nu\sigma}T^{\sigma\nu}+{\Gamma^\nu}_{\nu\sigma}T^{\mu\sigma}=0.
\een
Above the semicolon refers to the covariant derivative and ${\Gamma^\mu}_{\nu\sigma}$ to the Christoffel symbols.

From the knowledge of the expressions of the particle four-flow and energy momentum tensor components (\ref{tr14a}) -- (\ref{tr14f}) and the conservation laws (\ref{tr15}) one can obtain the field equations for the particle number density, momentum density and specific internal energy for a viscous and heat conducting fluid in the first post-Newtonian approximation. 

Here we are interested in determining the linearized field equations and for that end we consider a background state of constant values for the mass density, absolute temperature and Newtonian gravitational potential denoted by $\rho_0, T_0$ and $U_0$, respectively, which are superposed by linear perturbed fields denoted by $\rho_1, T_1, U_1, V_i^1, \Phi_1, \Pi_i^1$, namely
\ben\lb{tr16a}
\rho(\bx,t)=\rho_0+\rho_1(\bx,t),\qquad T(\bx,t)=T_0+T_1(\bx,t),\qquad U(\bx,t)=U_0+U_1(\bx,t),\\\lb{tr16b}
V_i(\bx,t)=V_i^1(\bx,t),\qquad \Phi(\bx,t)=\Phi_1(\bx,t),\qquad \Pi_i(\bx,t)=\Pi_i^1(\bx,t).
\een

 From the insertion of (\ref{tr14a}) into (\ref{tr15})$_1$ follows the linearized field equation for the mass density, by taking into account   the expressions of the Christoffel symbols in the first post-Newtonian approximation -- which can be found in \cite{Ch1,PGMK,GGKK} -- and of the representations (\ref{tr16a}), yielding
\ben\lb{tr17a}
\frac{\partial \rho_1}{\partial t}+\rho_0\frac{\partial V^1_i}{\partial x^i}+\frac{3\rho_0}{c^2}\frac{\partial U_1}{\partial t}=0,
\een

The linearized field equations  for the mass-energy and  momentum densities are obtained  from the time and spatial components  of (\ref{tr15})$_2$, respectively,  by considering the expressions  (\ref{tr14a}) -- (\ref{tr14f}), the representations (\ref{tr16a}), (\ref{tr16b}) and the  Christoffel symbols in the first post-Newtonian approximation. Hence it follows
\ben\lb{tr17b}
\frac{\partial \rho_1}{\partial t}+\rho_0\left(1+\frac{kT_0}{mc^2}\right)\frac{\partial V^1_i}{\partial x^i}+\frac{\rho_0}{c^2}\left(\frac{3kT_0}{2m}\frac{\partial T_1}{\partial t}+3\frac{\partial U_1}{\partial t}\right)-\frac{5k^2\rho_0T_0}{2m^2c^2\nu_0}\frac{\partial^2 T_1}{\partial x^i\partial x^i}=0,
\\\no
\rho_0\frac{\partial V^1_i}{\partial t}+\frac{k}m\left[1-\frac{1}{c^2}\left(\frac{5kT_0}{2m}+4U_0\right)\right]\left[T_0\frac{\partial \rho_1}{\partial x^i}+\rho_0\frac{\partial T_1}{\partial x^i}\right]-\rho_0\left[1-\frac{4U_0}{c^2}\right]\frac{\partial U_1}{\partial x^i}
\\\lb{tr17c}
-\frac{5k^2\rho_0T_0}{2m^2c^2\nu_0}\frac{\partial^2 T_1}{\partial t\partial x^i}-\frac{k\rho_0T_0}{m\nu_0}\left[1-\frac{3U_0}{c^2}\right]\left[\frac{\partial^2 V^1_i}{\partial x^j\partial x^j}+\frac13\frac{\partial^2 V^1_j}{\partial x^j\partial x^i}\right]-\frac{\rho_0}{c^2}\left[2\frac{\partial \Phi_1}{\partial x^i}+\frac{\partial\Pi^1_i}{\partial t}\right]=0.
\een

Since the constant values of the background state does not satisfy the Poisson equations  (\ref{tr4a}) and (\ref{tr4c}) it is usual to take into account the "Jeans swindle" (see e.g. \cite{Jeans,BT1,Coles}) which requires that the Poisson equations are valid only for  the  perturbed fields. Hence, by considering that $\varepsilon={3kT}/{2m}={3p}/{2\rho}$, the linearized Poisson equations become
\ben\lb{tr18a}
&&\nabla^2U_1=-4\pi G\rho_1,
\qquad
\nabla^2\Phi_1=-4\pi G\rho_1\left(U_0+\frac{9k}{4m}T_0\right)-4\pi G\rho_0\left(U_1+\frac{9k}{4m}T_1\right),
\\\lb{tr18b}
&&\nabla^2\Pi_i^1=-16\pi G\rho_0 V^1_i+\frac{\partial^2U_1}{\partial t\partial x^i}.
\een

Let us find a solution of the coupled system of partial differential equations (\ref{tr17a}) -- (\ref{tr18b}) in terms of a plane wave representation of the perturbed fields, namely
\ben\lb{tr19a}
\rho_1(\bx,t)=\overline{\rho}e^{\left[i\left(\kappa_ix^i-\omega t\right)\right]},\qquad T_1(\bx,t)=\overline{T}e^{\left[i\left(\kappa_ix^i-\omega t\right)\right]},\qquad U_1(\bx,t)=\overline{U}e^{\left[i\left(\kappa_ix^i-\omega t\right)\right]},
\\\lb{tr19b}
V_i^1(\bx,t)=\overline{V_i}e^{\left[i\left(\kappa_ix^i-\omega t\right)\right]},\qquad \Phi_1(\bx,t)=\overline{\Phi}e^{\left[i\left(\kappa_ix^i-\omega t\right)\right]},\quad \Pi_{i}^1(\bx,t)=\overline{\Pi}_{i}e^{\left[i\left(\kappa_ix^i-\omega t\right)\right]},
\een
where $\kappa_i$ denotes the wavenumber vector, $\omega$ the angular frequency and the  overlined quantities the small amplitudes of the wave.

We insert the plane wave representations (\ref{tr19a}) and (\ref{tr19b}) into the coupled system of partial differential equations (\ref{tr17a}) -- (\ref{tr18b}) and get a linearized system of algebraic equations for the amplitudes which reads 
\ben\lb{tr20a}
&&\omega_*\rho_*-V_*+\frac{3U_0}{c^2}U_*=0,
\\\lb{tr20b}
&&\omega_*\rho_*-\left(1+\frac{3c_s^2}{5c^2}\right)V_*+\frac{3U_0}{c^2}U_*+\frac{9c_s^2}{10c^2}\left(\omega_*+\frac{i\kappa_*}{\nu_*}\right)T_*=0,
\\\no
&&\left[\omega_*+\frac{4}{5\nu_*}\left(1-\frac{3U_0}{c^2}\right)i\kappa_*^2\right]V_*-\frac35\kappa_*^2\left[1-\frac{c_s^2}{c^2}\left(\frac{3}{2}+4\frac{U_0}{c_s^2}\right)\right]\left[\rho_*+T_*\right]
\\\lb{tr20c}
&&\qquad+\kappa_*^2\frac{U_0}{c_s^2}\left[1-\frac{4U_0}{c^2}\right]U_*-\frac{3c_s^2}{2c^2\nu_*}i\omega_*\kappa_*^2T_*+\frac{c_s^2}{c^2}\left(2\kappa_*^2\Phi_*-\omega_*  \Pi_*\right)=0,
\\\lb{tr20d}
&&\kappa_*^2\frac{U_0}{c_s^2} U_*= \rho_*,
\\\lb{tr20e}
&&\kappa_*^2\Phi_*=\left(\frac{U_0}{c_s^2}+\frac{27}{20}\right)\rho_*+\left(\frac{U_0}{c_s^2}U_*+\frac{27}{20}T_*\right),
\\\lb{tr20f}
&&\kappa_*^2 \Pi_*=4 V_*-\omega_*\kappa_*^2\frac{U_0}{c_s^2}U_*.
\een
Equations (\ref{tr20c}) and (\ref{tr20f}) result from the scalar product with $\kappa_i$.
Furthermore, the above equations were written in terms of the dimensionless quantities 
\ben
\kappa_i^*=\frac{\kappa_i}{\kappa_J},\qquad \omega_*=\frac\omega{\sqrt{4\pi G\rho_0}},\qquad \nu_*=\frac{\nu_0}{\sqrt{4\pi G\rho_0}},
\\\lb{tr21}
\rho_*=\frac{\overline\rho}{\rho_0},\qquad T_*=\frac{\overline T}{T_0},\qquad V_*=\frac{\overline V_i\kappa_i}{c_s\kappa_J},\qquad U_*=\frac{\overline U}{U_0},\qquad \Phi_*=\frac{\overline\Phi}{c_s^4},\qquad \Pi_*=\frac{\overline \Pi_i\kappa_i}{c_s^3\kappa_J},
\een
where  $\kappa_J=\sqrt{4\pi G\rho_0}/c_s$ denotes  the Jeans wavelength,  $c_s=\sqrt{5kT_0/3m}$ the sound speed  and $\kappa_*=\sqrt{\kappa_i^*\kappa_i^*}$.

The system of algebraic equations for the amplitudes (\ref{tr20a}) -- (\ref{tr20f}) admits a non-trivial solution if the determinant of the    coefficients which correspond to the amplitudes vanish. Hence it follows the dispersion relation which connect the dimensionless angular frequency $\omega_*$ with the dimensionless wavenumber $\kappa_*$, namely 
\ben\no
&&\omega_*^3+\frac{9i}{5\nu_*}\left\{\kappa_*^2+\frac43\left[1-\kappa_*^2\left(\frac5{12}+\frac{U_0}{c_s^2}\right)\frac{c_s^2}{c^2}\right]\right\}\omega_*^2+\bigg\{1-\kappa_*^2-\frac{4\kappa_*^4}{5\nu_*}+\bigg[\frac{33}{10}+\frac{2}{\kappa_*^2}
\\\no
&&\qquad+\frac{3\kappa_*^2}2-\frac{2U_0}{c_s^2}(1-2\kappa_*^2)-\frac{12\kappa_*^2}{5\nu_*^2}\bigg(1-\frac{U_0\kappa_*^2}{c_s^2}\bigg)\bigg]\frac{c_s^2}{c^2}\bigg\}\omega_*+\frac{i}{\nu_*}\bigg\{\kappa_*^2\bigg(1-\frac{3\kappa_*^2}5\bigg)
\\\lb{tr22a}
&&\qquad+\bigg[2+\frac{27\kappa_*^2}{10}\bigg(1+\frac{\kappa_*^2}3\bigg)-\frac{2\kappa_*^2U_0}{c_s^2}\bigg(1-\frac{6\kappa_*^2}5\bigg)\bigg]\frac{c_s^2}{c^2}\bigg\}=0.
\een
Here terms up to the order ${\cal O}(c^{-2})$ were taken into account.

In the case of a non relativistic and collisionless Boltzmann equation we have that $c_s/c\rightarrow0$ and $\nu_*\rightarrow\infty$ and
we obtain from (\ref{tr22a}) Jeans solution \cite{Jeans}
\ben\lb{tr22b}
\omega_*=\pm\sqrt{\frac{\lambda_J^2}{\lambda^2}-1}.
\een
Above we have introduced the wavelengths $\lambda$ and    $\lambda_J$ (Jeans wavelength)   through the relationship $\kappa_*=\kappa/\kappa_J=\lambda_J/\lambda$. In the case of small wavelengths with respect to Jeans wavelength $\lambda_J/\lambda>1$ the dimensionless angular frequency is a real quantity and the perturbations propagate as harmonic waves in time. On the other hand, for big wavelengths $\lambda_J/\lambda<1$ the angular frequency becomes a pure imaginary quantity and the perturbations will grow or decay in time, which will depend on the sign of the solution (\ref{tr22b}). The perturbations which grow in time are referred  as Jeans instability, which is associated with the gravitational collapse of self-gravitating gas clouds.

The analysis of Jeans instability within the first and second post-Newtonian approximation by considering the Eulerian hydrodynamic equations were investigated in \cite{NKR,NH,gg1} and \cite{gg2}, respectively. Here if we consider a collisionless Boltzmann equation where   $\nu_*\rightarrow\infty$  (\ref{tr22a}) reduces to
\ben\lb{tr22c}
\omega_*^3+\bigg\{1-\kappa_*^2+\bigg[\frac{33}{10}+\frac{2}{\kappa_*^2}
+\frac{3\kappa_*^2}2-\frac{2U_0}{c_s^2}(1-2\kappa_*^2)\bigg]\frac{c_s^2}{c^2}\bigg\}\omega_*=0,
\een
which is the dispersion relation in the first post-Newtonian approximation where dissipative effects are not considered. There is a  difference of this expression with the one in \cite{GGKK}, since here the constant value  is $33/10$ while there is $9/2$.  The reason of this difference is that here we have considered the mass, mass-energy and momentum densities hydrodynamic equations while in the former work only the mass and momentum densities hydrodynamic equations were taken into account.

For big wavelengths with respect to Jeans wavelength $\lambda_J/\lambda<1$ three different values associated with the dimensionless angular  frequencies can be obtained from (\ref{tr22a}) which correspond to the growth/decay of the perturbations: 
\ben\lb{tr23a}
&&\omega_*=-\frac{i}{\nu_*}\frac{\lambda_J^2}{\lambda^2}\left[1-\frac{7c_s^2}{5c^2}\right]+\dots,
\\\lb{tr23b}
&&\omega_*=i\left[1-\frac12\frac{\lambda_J^2}{\lambda^2}\left(1+\frac4{5\nu_*}\right)+\left(\frac{43}{20}-\frac{U_0}{c_s^2}+\frac{\lambda^2}{\lambda_J^2}-\frac6{5\nu_*}\right)\frac{c_s^2}{c^2}\right]+\dots,
\\\lb{tr23c}
&&\omega_*=-i\left[1-\frac12\frac{\lambda_J^2}{\lambda^2}\left(1-\frac4{5\nu_*}\right)+\left(\frac{43}{20}-\frac{U_0}{c_s^2}+\frac{\lambda^2}{\lambda_J^2}+\frac6{5\nu_*}\right)\frac{c_s^2}{c^2}\right]+\dots\,.
\een

On the other hand, if we expand the dimensionless wavenumber in power series of the reduced angular frequency $\kappa_*=a_0+a_1\omega_*+\dots$ we get from the dispersion relation (\ref{tr22a}) the solution where the perturbations propagate as harmonic waves
\be
\kappa_*=\sqrt{\frac53}\left[1+\left(\frac{27}{10}+U\right)\frac{c_s^2}{c^2}\right]+\frac{2i}{3\nu_*}\sqrt{\frac53}\left\{1+\frac{3\nu_*^2}{10}+\left[\frac{24}5+2U-\nu_*^2\left(\frac{3U}{10}+\frac{36}{25}\right)\right]\frac{c_s^2}{c^2}\right\}\omega_*+\dots\,.
\ee{tr23d}

\section{Constitutive equations}\lb{sec5}

As was previously said the thermodynamic theory of a  single relativistic fluid is characterized by the fields of particle four-flow $N^\mu$ and energy-momentum tensor $T^{\mu\nu}$ whose   
hydrodynamic equations are  the conservation laws (\ref{tr15}).

The representation of the particle four-flow and energy-momentum tensor in terms of non-relativistic quantities makes use of the four-velocity $U^\mu$ --where $U^\mu U_\mu=c^2$ -- and of the projector $\Delta^{\mu\nu}=g^{\mu\nu}-U^\mu U^\nu/c^2$ -- where $g^{\mu\nu}$ denotes the metric tensor. The projector has the properties $\Delta^{\mu\nu}U_\nu=0$, $\Delta^{\mu\nu}\Delta_{\nu\sigma}={\Delta^\mu}_\sigma$ and in a local Minkowski rest frame where $U^\mu=(c,\bf0)$ it reduces to $\Delta^{\mu\nu}={\rm diag}(0,-1,-1,-1)$.

Two representations for  the particle four-flow and energy-momentum tensor  in terms of non-relativistic quantities are the Eckart \cite{Eck} and the Landau-Lifshitz \cite{LL} decompositions. Here we shall use the Eckart decomposition where  the particle four-flow and energy-momentum tensor  are written as
\ben\lb{tr1a}
&&N^{\mu}=nU^{\mu},
\\\lb{tr1b}
&&T^{\mu\nu}=p^{\langle \mu\nu\rangle}-\left(p+\varpi\right)
\Delta^{\mu\nu}
+\frac\epsilon{c^2}
U^{\mu}U^{\nu}
+\frac1{c^2}
\bigg(U^{\mu}q^{(\nu)}
+U^{\nu}q^{(\mu)}\bigg).
\een
Above $n$ is the particle number density, $p$ the hydrostatic pressure, $\varpi$ the non-equilibrium pressure, 
$p^{\langle \mu\nu\rangle}$ the pressure deviator,
 $q^{(\mu)}$ the heat flux and $\epsilon$ the energy density. The energy density  is a sum of two terms one related with the internal energy density $\rho\varepsilon$  while the other with the mass density $\rho$, namely $\epsilon=\rho c^2(1+\varepsilon/c^2)$. The following projections of the particle four-flow and energy-momentum tensor define the non-relativistic quantities (see e.g \cite{CK}):
\ben\lb{tr2a}
&&n=\frac1{c^2} N^{\mu} U_{\mu}, \qquad
\epsilon=\frac1{c^2} U_{\mu}
T^{\mu\nu}U_{\nu},\qquad \left (p+\varpi\right)
=-\frac13 \Delta_{\mu\nu}T^{\mu\nu}
\\\lb{tr2b}
&&p^{\langle \mu\nu\rangle}
=\left(
{\Delta^{\mu}}_{\sigma}
{\Delta^{\nu}}_{\tau}
-\frac13
\Delta^{\mu\nu}\Delta_{\sigma\tau}\right)
T^{\sigma\tau},
\qquad
q^{(\mu)}={\Delta^{\mu}}_{\nu}U_{\sigma}
T^{\nu\sigma},
\een

In the first post-Newtonian approximation the components of the four-velocity read \cite{Ch1,Wein,GGKK}
\ben\lb{tr5a}
&&U^0=c\bigg[1+\frac1{c^2}\bigg(\frac{V^2}2+U\bigg)\bigg],
\qquad U^i=\frac{V_iU^0}c,
\een
where $\bf V$ denotes the hydrodynamic three velocity.

From the knowledge of the components of the metric tensor in the first post-Newtonian approximation 
\ben
g_{00}=1-\frac{2U}{c^2}+\frac2{c^4}\left(U^2-2\Phi\right),\qquad g_{0i}=\frac{\Pi_i}{c^3},\qquad g_{ij}=-\left(1+\frac{2U}{c^2}\right)\delta_{ij},
\een
and of the four-velocity components (\ref{tr5a}) we can determine the components of the projector, which read
\ben\lb{tr6a}
&&\Delta^{00}=-\frac{V^2}{c^2}-\frac1{c^4}\left(6UV^2+V^4-2\Pi_iV_i\right),\qquad
\Delta^{0i}=-\frac{V_i}{c}-\frac1{c^3}\left(2UV_i+V^2V_i-\Pi_i\right),\\\lb{tr6b}
&&\Delta^{ij}=-\left(1-\frac{2U}{c^2}\right)\delta_{ij}-\frac{V_iV_j}{c^2}.
\een

Now we introduce the  non-relativistic pressure deviator 
\ben
\mathfrak{p}_{ij}=p_{ij}-{p_{kk}}\delta_{ij}/3 \qquad\hbox{whit } \qquad\delta_{ij}\mathfrak{p}_{ij}=0,
\een
so that the components of the pressure deviator $p^{\langle \mu\nu\rangle}$  become \cite{GKN}
\ben\lb{tr7a}
&&p^{\langle ij\rangle}=\mathfrak{p}_{ij}+\frac1{2c^2}\left(\mathfrak{p}_{ik}V_kV_j+\mathfrak{p}_{jk}V_kV_i\right),
\\\lb{tr7b}
&&p^{\langle00\rangle}=\mathfrak{p}_{ij}\frac{V_iV_j}{c^2},
\qquad
p^{\langle0i\rangle}=\mathfrak{p}_{ij}\frac{V_j}c.
\een

In terms of the non-relativistic heat flux vector $\mathfrak{q}_i$  the  components of  the heat flux $q^{(\mu)}$ are
 \ben\lb{tr7d}
q^{(i)}=\mathfrak{q}_i, \qquad  q^{(0)}=\mathfrak{q}_i\frac{V_i}c.
\een

In the five field thermodynamic theory -- where the basic fields are the mass density, momentum density and internal energy density -- the  pressure deviator, the dynamic pressure and  the heat flux vector are given by constitutive equations. Here we can obtain the desired constitutive equations from the components of the energy-momentum tensor (\ref{tr14a}) -- (\ref{tr14f}) combined with the decomposition expressions  (\ref{tr2a}) and (\ref{tr2b}) and the components of the projection (\ref{tr6a}) and (\ref{tr6b}).
Hence it follows the constitutive equations for the non-relativistic heat flux vector and pressure deviator
\ben\no
&&\mathfrak{q}_i=\underline{-\frac{5kp}{2m\nu}\bigg(1-\frac{c_s^2}{c^2}\frac{U}{c_s^2}\bigg)\frac{\partial T}{\partial x^i}}+\frac{p}{\nu c^2}\Delta_{ijkl}\frac{\partial V_k}{\partial x^l}\bigg[\bigg(\frac{5kT}{2m}+3U+\frac{V^2}2\bigg)V_j-\Pi_j\bigg]
\\\lb{tr25a}
&&+\frac{p}{\nu c^2}(V^2\delta_{ij}-V_iV_j)\bigg[V_k\frac{\partial V_k}{\partial x^j}-\frac{\partial T}{\partial x^j}\bigg]
+\frac{p}{\nu c^2}\bigg(V^2\delta_{ij}+\frac{V_iV_j}3\bigg)\bigg(\frac{\partial U}{\partial x^j}-\frac1\rho\frac{\partial p}{\partial x^j}\bigg),
\\\no
&&\mathfrak{p}_{ij}=\underline{-\frac{p}\nu\bigg[1+\frac{c_s^2}{c^2}\bigg(\frac{3}{2}-\frac{U}{c_s^2}\bigg)\bigg]\Delta_{ijkl}\frac{\partial V_k}{\partial x^l}}
+\frac{2p}{3\nu c^2}\frac{\partial V_k }{\partial x^k}\bigg(V_iV_j-\frac13V^2\delta_{ij}\bigg)
\\\lb{tr25b}
&&-\frac{p}{\nu c^2}\Delta_{ijkl}\left[\frac12\frac{\partial V^2V_k}{\partial x^l}+V_k\left(\frac{\partial U}{\partial x^l}-\frac1\rho\frac{\partial p}{\partial x^l}\right)\right].
\een

The constitutive equation for the dynamic pressure $\varpi$ does not show up in the first post-Newtonian approximation and it is known  that in the kinetic theory of relativistic gases the coefficient of bulk viscosity -- which relates the dynamic pressure with the velocity divergent -- is of order $\mathcal{O}(c^{-4})$ (see e.g. \cite{CK}).  

Let us fix our attention in the underlined linearized terms  in (\ref{tr25a}) and (\ref{tr25b}). Without the relativistic corrections they reduce to the non-relativistic constitutive equations of a viscous and heat conducting gas, namely
\ben
\mathfrak{q}_i=-\frac{5kp}{2m\nu}\frac{\partial T}{\partial x^i},\qquad \mathfrak{p}_{ij}=-\frac{p}\nu\Delta_{ijkl}\frac{\partial V_k}{\partial x^l},
\een
where the thermal conductivity $\lambda$ and the shear viscosity $\mu$ coefficients are those of the non-relativistic BGK model
\ben
\lambda=\frac{5kp}{2m\nu},\qquad \mu=\frac{p}\nu.
\een
With the first post-Newtonian correction these coefficients read
\ben
\lambda=\frac{5kp}{2m\nu}\bigg(1-\frac{c_s^2}{c^2}\frac{U}{c_s^2}\bigg),\qquad \mu=\frac{p}\nu\bigg[1+\frac{c_s^2}{c^2}\bigg(\frac{3}{2}-\frac{U}{c_s^2}\bigg)\bigg].
\een
We note that the coefficients of shear viscosity and thermal conductivity do depend on the Newtonian gravitational potential.  On the basis of a non-relativistic kinetic theory the influence the gravity on the thermal coefficient was first reported  in \cite{And1,And2}. Within the framework of a relativistic kinetic theory the transport coefficients of shear viscosity, thermal conductivity and bulk viscosity were obtained by considering a Schwarzschild metric in \cite{sch1} and the diffusion coefficient in \cite{sch2}.%By considering the non-relativistic limiting case and a weak gravitational field  the expressions for the thermal conductivity  and shear viscosity coefficients read \cite{sch1}:\ben\lambda=\frac{5kp}{2m\nu}\bigg[1+\frac{c_s^2}{c^2}\bigg(\frac52-\frac{2U}{c_s^2}\bigg)\bigg],\qquad \mu=\frac{p}\nu\bigg[1+\frac{c_s^2}{c^2}\bigg(\frac{3}{2}-\frac{4U}{c_s^2}\bigg)\bigg].\een

\section{Conclusions}\lb{sec6}
In this work we have examined a relaxation-time model for the post-Newtonian Boltzmann equation and determined the non-equilibrium distribution function by using the Chapman-Enskog method and the equilibrium post-Newtonian Maxwell-J\"uttner distribution function. The components of the energy-momentum tensor were calculated by using the equilibrium and non-equilibrium distribution functions. From the conservation laws of the particle four-flow and energy-momentum tensor the linearized field equations for the mass, momentum and internal energy densities were determined. A plane wave solution of these linearized field equations coupled with the three post-Newtonian Poisson equations was found. By using the Eckart decomposition of the energy-momentum tensor the constitutive equations for  the viscous stress and heat flux vector were obtained and it was shown that the transport coefficients of  shear viscosity and heat conductivity do  depend on the Newtonian gravitational potential.
\acknowledgments{ This work was supported by Conselho Nacional de Desenvolvimento Cient\'{i}fico e Tecnol\'{o}gico (CNPq), grant No.  304054/2019-4.}

\end{document}